\documentclass[%
 aip,
 amsmath,amssymb,
 reprint,%
]{revtex4-2}
\usepackage{graphicx}
\usepackage{dcolumn}
\usepackage{bm}

\usepackage[utf8]{inputenc}
\usepackage[T1]{fontenc}
\usepackage{mathptmx}
\usepackage{etoolbox}

\makeatletter
\def\@email#1#2{%
 \endgroup
 \patchcmd{\titleblock@produce}
  {\frontmatter@RRAPformat}
  {\frontmatter@RRAPformat{\produce@RRAP{*#1\href{mailto:#2}{#2}}}\frontmatter@RRAPformat}
  {}{}
}%
\makeatother
\begin{document}
                                                                                                                                                                                                                                                                                                                                                                                                                                                                                                                                                                                                                                                                                                                                                                                                                                                                                                                                                                                                                                                                                                                                                                                                                                                                                                                                                                                                                                                                                                                                                                                                                                                                                                                                                                                                                                                                                                                                                     
\title{Detection of terahertz radiation using  topological graphene micro- nanoribbon  structures with  transverse plasmonic resonant cavities
}
\author{V.~Ryzhii$^{1*}$,   C.~Tang$^{1,2}$,  T. Otsuji$^{1}$,  M.~Ryzhii$^{3}$,  and M. S. Shur$^4$}
\address{
$^1$Research Institute of Electrical Communication,~Tohoku University,~Sendai~ 980-8577,
Japan\\
$^2$Frontier Research Institute for Interdisciplinary Sciences,
Tohoku University, Sendai 980-8578, Japan\\
$^3$School of Computer Science and Engineering, University of Aizu, Aizu-Wakamatsu 965-8580, Japan\\
$^4$Department of Electrical, Computer, and Systems Engineering,\\ Rensselaer Polytechnic Institute,~Troy,~New York~12180,\\ USA\\
*{Author to whom correspondence should be addressed: v-ryzhii@gmail.com}
}
\begin{abstract}
  The lateral interdigital array of the  graphene  microribbons (GMRs) on the h-BN substrate connected by narrow graphene nanoribbon (GNR) bridges serves as an
  efficient detector of terahertz (THz) radiation.
  The detection is enabled by the  nonlinear GNR 
elements providing the rectification of the THz signals. 
 The excitation of plasmonic waves along the GMRs (transverse plasmonic oscillations) 
by impinging THz radiation can  
 lead to a strong resonant 
amplification of the rectified signal current and substantial enhancement of the detector response. The  GMR arrays with the  GNR bridges  can be formed by the perforation of uniform graphene layers.
\end{abstract}


\maketitle

\section{Introduction}

\begin{figure}[t]
\centering
\includegraphics[width=7.0cm]{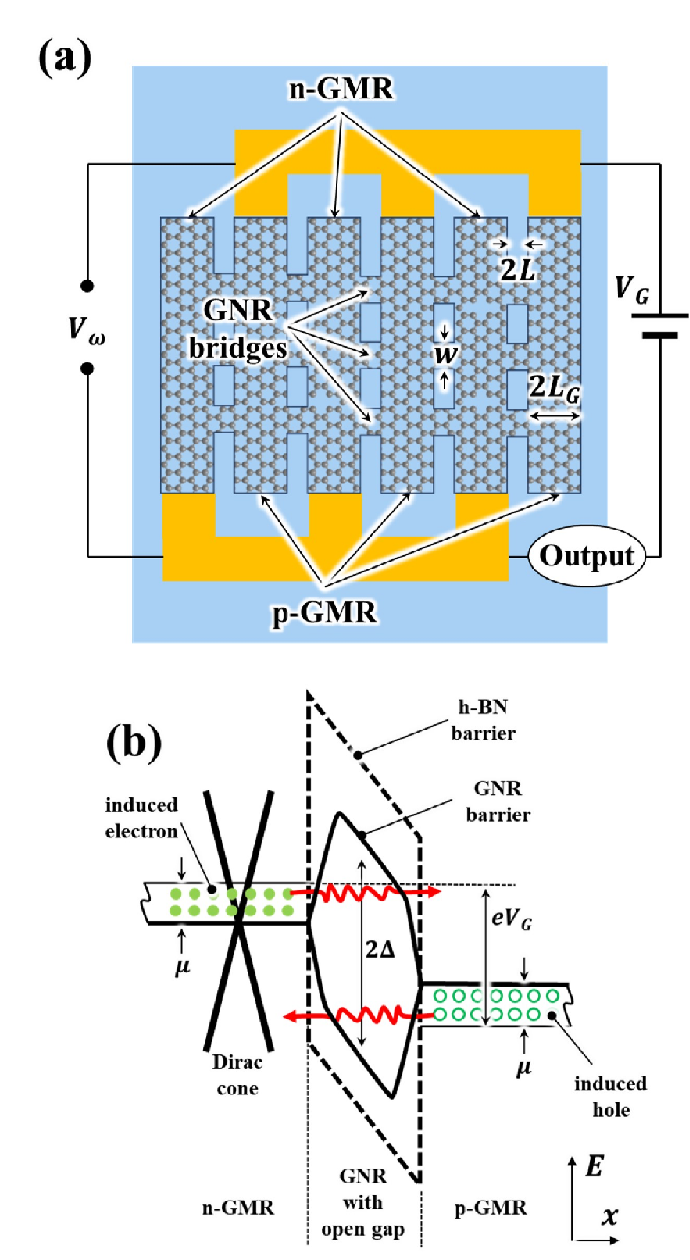}
\caption{Schematic top view of (a) the GMR-GNR THz detector  and (b) band diagram (one period) across the GNRs ($z = z_n$, solid lines) and across perforations
(dashed lines)} \label{Fig1}
\end{figure}
The pioneering work by Dyakonov and Shur [1] has stimulated
extended theoretical and experimental studies of plasmonic effects
in two-dimensional (2D) and one-dimensional (1D) heterostructures.
This has resulted in the proposals and  realization of  different devices, including the terahertz 
(THz) detectors and sources using the plasmonic resonances based on the field-effect transistors (FETs) and the kindred heterostructures (see, for example, Refs.~2 and 3), including those with the graphene layer channels. \cite{4,5,6,7,8,9,10,11,12,13,14,15}
 In the FET-like devices with the gated electron or hole channels,  
 the plasmons are associated with a distributed inductance due to
  the inertia of carrier motion along channels,~\cite{1,16} and with the distributed 
the  channel-gate capacitance.~\cite{1,5,17} As predicted recently,~\cite{18,19} 
  the ungated coplanar graphene microribbon (GMR) structures can exhibit
  the plasmonic response, with the resonant plasmonic frequency 
    determined by the inter-GMR capacitance. This capacitance depends on the spacing between the neighboring GMRs and their width.~\cite{20,21,22}

In this paper, we propose the concept of the THz radiation detector based on a coplanar interdigital GMR array, in which the neighboring GMRs are connected by narrow graphene nanoribbons (GNRs).

We develop the device model, and  calculate the detector responsivity and the noise equivalent power. 
We show that the optimized GMR-GNR detectors under consideration can exhibit  elevated performance.

\section{GMR-GNR detector device structure}

Figure~1(a)  schematically shows the topological structure of the  GMR-GNR  array of the $M$ pairs of the  interdigital  GMRs. The topological properties of this device are due to two-dimensionality of graphene they could be bended if deposited on a flexible substrate.

Each GMR pair is connected by $(2N-1)$ GNRs bridges, where  $N = 1,2,3,...$. The GMR-GNR array is placed on the substrate, which does not degrade substantially the carrier mobility in the GMRs (for example, on the h-BN substrate).   The simplest GMR-GNR detector structure may have just two GMRs (connected by one GNR). 

The bias voltage between the GMR ends, $V_G$, enables the formation of the 2D  electron and hole gases. 
The  bias voltage 
is split between the array  dc resistance $r_{GNR}$ and the load resistance $r_L$ (if any, depending on the method of signal output). 
Thus,
 the coplanar GMR-GNR detector structure  constitutes a lateral periodic p-GMR/n-GMR/p-GMR/…/n-GMR array. The gap opening in the GNRs leads to the formation of the energy barrier for the electron in the n-GMR and holes in the p-GMR. The operation of the  GMR-GNR  detector is associated with the excitation of the standing plasmonic waves along the GMRs by the signal ac voltage between their ends, $V_{\omega}(t)$, 
induced by the impinging THz radiation collected by an antenna. 
 These voltages result in the AC currents through the GNR  bridges amplified by the rectified plasmonic resonant response. The rectified response can support the output current or output voltage  signals.
It is assumed that the input circuit incorporates a DC block to isolate the input and output circuits.

The GMRs have a length $2H$ and a  width $2L_G$.  The GNR transverse characteristic width and the GNR length are  $w$ and  $2L$, respectively [see Figs.~1(a) and 1(b)]. 
  We assume that $H \gg L_G, L$ and $2H \gg (2N-1)w$.
The latter implies that the spacing between the GNRs $D \simeq 2H/(2N-1)$  is sufficiently large to prevent the carrier interaction in the neighboring GNRs and capacitive  coupling between  GNRs.

 One of the GMR-GNR detector features is that  the plasmonic wave-vectors directed along the GMRs are transverse to the inter-GMR current.
  The  topological GMR-GNR detector structures  can be, in particular, formed by the uniform graphene layer perforation.
Similar devices can be based on the interdigital GMR arrays using
semiconducting carbon nanotubes as bridges connecting GMRs.

Figure 1(b)  shows the band diagram at the structure $z$-cross sections 
In the  perforations (i.e., between the GNRs), the n- and p-GMRs are separated by relatively high energy barriers
for the electrons and holes [see the dashed line in Fig.~1(b)]. 
The heights of these barriers are determined by the band alignment
between graphene and  the substrate material, for example, h-BN
(although other substrate materials providing sufficiently high-quality interface and, hence, high carrier mobility in the GMRs, in particular SiC, can be used). 
In contrast, the GNRs provide relatively low energy barriers  between the n- and p-GMRs [as shown by the  solid line in Fig.~1(b)] allowing effective tunneling. These barriers are associated with the lateral confinement  of  the electron and hole motion
(perpendicular to the GNRs) and the pertinent quantization of their
energy spectra. 
Hence,   the height of such barriers is determined by the GNR thickness and doping.
As a result, the DC and AC electron and hole currents between the neighboring GMRs
flow through the GNRs. 
We consider the arrays, in which the GNRs width is close to the characteristic value $w$ (estimated below), except, possibly, the small transition regions near the GMR-GNR contacts.
The energy barrier shape in such arrays,is virtually trapezoidal turning to a triangular one at sufficiently strong bias voltage $V_G$.   

The linear and nonlinear components of the inter-GMR  currents across the GNRs $j$ are determined by
the  tunneling processes. The tunneling currents  are characterized by the differential conductance $\sigma_{GNR} =(dJ/dV)|_{V_G}$ and the  current-voltage nonlinearity parameter $\eta_{GNR} = \frac{1}{2}(d^2J/dV^2)|_{V_G}$. Here $J= J(V)$ is the
inter-GNR current between the GMR edges via one GNR. 

The impinging THz radiation received by an antenna induces the signal voltage $V_{\omega}(t) = V_{\omega}\exp(-i\omega t)$, where $V_{\omega} \propto \sqrt{P_{\omega}}$  is  the THz signal voltage amplitude,  $\omega$ is the frequency, 
 and $P_{\omega}$ is the THz radiation power collected by the device antenna.
The inter-GMR AC displacement current  is distributed along the GMRs. It is characterized by
the inter-GMR capacitance, $c_G$, per unit of the GMR length.

Since  all the GMR-GNR array periods are equivalent, the net signal current density $j_{\omega} = j_{\omega}(z)$
 (current per GMR unit length)  between the neighboring GMRs (including the electron and hole currents via the GNRs and the displacement current) can be presented as

\begin{eqnarray}\label{eq1}
 j_{\omega} = -i\omega\,c_G( \varphi_{\omega}^+ - \varphi_{\omega}^-)\nonumber\\
 + \sum_{n=-N}^{N}[\sigma_{GNR}( \varphi_{\omega}^+ - \varphi_{\omega}^-)
+ \eta_{GNR}( \varphi_{\omega}^+ - \varphi_{\omega}^-)^2]\,\delta(z -  z_n).
\end{eqnarray} 
Here $\varphi_{\omega}^+ = \varphi_{\omega}^+(z)$ and  $\varphi_{\omega}^-= \varphi_{\omega}^-(z)$ are the signal components of the p- and n-GMRs
(depending on the coordinate $z$ directed along the GMRs), respectively, $z_n$ is the $z$-coordinate of 
the  $n$-th GNR, and $\delta(z-z_n)$ is the form-factor characterizing the z-distribution of the carrier current through the GNR, which, due to the narrowness of the GNRs,  is replaced by
the Dirac delta function [with $\int dz \delta(z-z_n) = 1$].

Considering the specifics of the GMR shape (blade-like), the inter-GMR capacitance
per unit length  is presented as
$c_G= (\kappa/2\pi^2){\overline c}_G$~\cite{20}  with
 ${\overline c}_G  = a\tan^{-1}\biggl(\frac{1}{\sqrt{a^2-1}}\biggr) 
+\ln(a+\sqrt{a^2-1})$, $a = L_G/L$, and $\kappa = (\kappa_S+1)/2$, where $\kappa_S$ is the dielectric constant of the substrate. 
We   disregarded the transit delay of the electrons and holes in assuming that the GNR  length $2L$ is sufficiently small.

Considering the inter-GMR currents via the GNRs  in the right-hand side of Eq.~(1) as a small perturbation and accounting for the balance between the carriers induced in the GNRs and their output/input at the side contacts, we have the following equation for the linear components of $\varphi_{\omega}^+$ and  $\varphi_{\omega}^-$ :
\begin{eqnarray}\label{eq2}
 \sigma_{G,\omega}\frac{\partial^2 \varphi_{\omega}^{\pm}}{\partial z^2} = 
 \mp\,i\omega\,c_G( \varphi_{\omega}^+ - \varphi_{\omega}^-).
\end{eqnarray}
Here 
$\sigma_{G,\omega} =\sigma_{G}[i\nu/(\omega +i\nu)]$ is the AC Drude longitudinal conductance of the GMRs (of the width equal to $2L_G$) with
$\sigma_{G} = (2e^2\mu\,L_{G}/\pi\hbar^2\nu)$ being its DC value, and $\mu$ is the electron and hole Fermi energy
and $\nu$ is   the electron and hole collision frequency in the GMRs.
The Fermi energy can be expressed via the steady-state carrier density in the GNRs $\Sigma_G$: 
$\mu \simeq \hbar\,v_W\sqrt{\pi\Sigma_G}$, 
where $v_W \simeq 10^8$~cm/s is the characteristic electron and hole velocity in graphene,  and $\hbar$ is the Planck constant. 
Accounting for that $\Sigma_G \simeq c_GV_G/2eL_G$, for the Fermi energy of the carriers electrically induced carriers in the GMRs, we obtain 
$\mu \simeq \hbar\,v_W\sqrt{c_GV_G/2eL_G} =\sqrt{{\overline V}_GV_G}$ and ${\overline V}_G =(\pi\,c_G\hbar^2v_W^2/2e^3L_G)$.

The boundary conditions given at the ends of the GMRs are: 

\begin{eqnarray}\label{eq3}
 \varphi_{\omega}^{\pm}|_{z = \pm H} = \pm V_{\omega}/2,\qquad
\frac{\partial  \varphi_{\omega}^{\pm}}{\partial z}\biggr|_{z\mp H} =0.
\end{eqnarray} 

\section{Resonant excitation of plasmonic oscillations by impinging THz radiation}

Using Eqs.~(2) and (3),  we obtain

\begin{eqnarray}\label{eq4}
\varphi_{\omega}^{+} = \frac{\cos(\gamma_{\omega}z/H)}{(\cos\gamma_{\omega} -\gamma_{\omega} \sin\gamma_{\omega})} \frac{V_{\omega}}{2},
\end{eqnarray}

\begin{eqnarray}\label{eq5}
\varphi_{\omega}^{-} =  -\frac{\cos(\gamma_{\omega}z/H)}{(\cos\gamma_{\omega} -\gamma_{\omega} \sin\gamma_{\omega})} \frac{V_{\omega}}{2}. 
\end{eqnarray}
Here

\begin{eqnarray}\label{eq6}
 \gamma_{\omega} =
 \frac{H\sqrt{2\omega(\omega+i\nu)}}{S} = 
 \frac{ \pi\sqrt{\omega(\omega+i\nu)}}{2\Omega},
 \end{eqnarray} 
where 

\begin{eqnarray}\label{eq7}
S = \sqrt{\frac{2e^2\mu\,L_G}{\pi\,c_G\,\hbar^2}}, \qquad
\Omega = \frac{e}{H\,\hbar}\sqrt{\frac{  \pi\mu\,L_G}{4c_G}}
\end{eqnarray}
are the characteristic velocity of the plasmonic wave along the GNRs and their
  characteristic frequency (plasmonic frequency) associated with the inter-GNR capacitance and the carrier inductance (defined as in Ref.~13), respectively, and $\nu$ is the frequency  of the carrier 
  scattering on impurities and acoustic phonons in the GMRs. Equations~(4) and (5) yield the following spatial distribution of the ac voltage swing between the p- and n-GNRs:

\begin{eqnarray}\label{eq8}
\varphi_{\omega}^{+} - \varphi_{\omega}^{-} =  \frac{\cos(\gamma_{\omega}z/H)}{(\cos\gamma_{\omega} -\gamma_{\omega} \sin\gamma_{\omega})} V_{\omega}.
\end{eqnarray}

\section{Rectified signal current and voltage}

Accounting for Eqs.
~(1) and (8), for the rectified signal current, $\Delta J_{\omega}$ (averaged over the period $2\pi/\omega$), 
we arrive at

\begin{eqnarray}\label{eq9}
\Delta J_{\omega}  = \frac{M\,\eta_{GNR}}{2}\int_{-H}^H dz\sum_{n=1}^{} | \varphi_{\omega}^+ - \varphi_{\omega}^-|^2 \, \delta (z-z_n).
\end{eqnarray} 

Assuming that  the number of the GNRs $(2N-1) \gg 1$,
from Eqs.~(8) and (9) we obtain

\begin{eqnarray}\label{eq10}
\Delta J_{\omega} \simeq \frac{M(2N-1)}{2}\,\eta_{GNR}\Pi_{\omega}V_{\omega}^2
\end{eqnarray} 
with

\begin{eqnarray}\label{eq11}
\Pi_{\omega} =  
\frac{|1+\sin \gamma_{\omega}\cos\gamma_{\omega}|}
{|\cos\gamma_{\omega} -\gamma_{\omega} \sin\gamma_{\omega}|^2} 
.
\end{eqnarray}

The frequency-dependent quantity $ \Pi_{\omega}$
is  the main factor describing the plasmonic response of the GMR array.

In the circuit with the load resistance,   the rectified voltage across the load resistance $\Delta V_{\omega}$ is presented as

\begin{eqnarray}\label{eq12}
\Delta V_{\omega}  = \frac{M(2N-1)r_L\rho_{GNR}}{(r_L + \rho_{GNR})}\eta_{GNR}\Pi_{\omega}
V_{\omega}^2.
\end{eqnarray}
Here $\rho_{GNR} = [M(2N-1)\sigma_{GNR}]^{-1}$ is the GMR-GNR array differential resistance [see Eq.~(A11)].

\section{Responsivity}
Coupling between the impinging THz radiation and the plasmonic oscillations in the GMRs is given by the antenna, so that, in principle, there is no right relation between
the THz radiation polarization and the plasmon wave-vector. Due to this, the role of the  antenna reduced to the generation of the AC voltage $V_{\omega}$ as shown in Fig.~1(a).
Considering  
that $V_{\omega}$ and the THz power $P_{\omega}$ are related as $V_{\omega}^2 = 16\pi^2P_{\omega}/c$, where $c$ is the speed of light in vacuum and the antenna gain is set $g=2$, 
 from Eqs.~(10) and (12)  we arrive at the following expressions for the GMR-GNR detector current  responsivity,  $R_{\omega}^J = \Delta J_{\omega}/P_{\omega}$ 
(in the A/W units)

\begin{eqnarray}\label{eq13}
 R_{\omega}^J =
 R^J\Pi_{\omega}
\end{eqnarray}
with
 
\begin{eqnarray}\label{eq14}
 R^J = 8\pi^2M(2N-1)\frac{\eta_{GNR}}{c},
\end{eqnarray}
 being the characteristic current  responsivity of the GMR-GNR detectors under consideration.
 Using Eq.~(14) and invoking the pertinent relations  from the Appendix,
 we arrive at the following formula for the
  characteristic responsivity:

\begin{eqnarray}\label{eq15}
 R^J \simeq 16\pi\,M(2N-1)\frac{e^2}{c\hbar}
\sqrt{\frac{{
\overline V}_G}{V_G}}\frac{V_{tunn}^2}{V_G^3}\exp\biggl(-\frac{V_{tunn}}{V_G}\biggr)
\nonumber\\ 
=\frac{16\pi\,M(2N-1)}{137}
\sqrt{\frac{{
\overline V}_G}{V_G}}\frac{V_{tunn}^2}{V_G^3}\exp\biggl(-\frac{V_{tunn}}{V_G}\biggr).
\end{eqnarray}
Here 
\begin{eqnarray}\label{eq16}
{\overline V}_G =\frac{\pi\,c_G\hbar^2v_W^2}{2e^3L_G}
\end{eqnarray}
and

\begin{eqnarray}\label{eq17}
V_{Tunn} = \frac{\pi\Delta^2L}{\hbar\,v_We}= \frac{\pi^3\hbar\,v_WL}{ew^2}.
\end{eqnarray}
are 
the voltage characterizing the sensitivity of the carrier density in the GMRs to the bias voltage
and the  tunneling voltage, respectively, and $W$ is the effective GNR width ($W \gtrsim w)$.

As follows from Eq.~(15), the characteristic current responsivity of the GMR-GNR detectors $R^V$ is a nonmonotonic function of the bias voltage $V_G$. It reaches a maximum at $V_{Max} = 2V_{Tunn}/7$, with

\begin{eqnarray}\label{eq18}
{\rm max}~ R^J \simeq 
0.889\,M(2N-1)\frac{\sqrt{{
\overline V}_G}}{V_{Tunn}^{3/2}}.
\end{eqnarray}

\begin{table*}[t]
\centering 
\vspace{2 mm}
\begin{tabular}{|r|c|c|c|c|c|c|c|c|c|c|c|}
\hline
\hline
Structure& $H$  (nm)&  $L_G$ (nm) & $L$ (nm) &$W/w$ (nm)& $\Delta$(meV)& $c_G$& 
${\overline V}_G$~(mV) &$V_{Tunn}$ (mV)& $V_{Max}$ (mV) &$V_{Min}$~(mV)& 
$\nu$ (ps$^{-1}$)\\ 
\hline
I &	250,\, 375,\, 500&	20& 10 &	12.5/9.5&157&	    0.657& 14  &1238& 354 &165& 2,\, 4,\, 6\\
\hline
II&250,\, 375,\,	500 & 	15& 10&	10.0/9.5 &	196&	0.576& 16  & 1935& 553 & 258 &2,\, 4,\, 6\\ 
\hline
III &250,\, 375,\,	500&	10& 7.5	 & 7.5/5/5&	261&	    0.438& 19 &2580& 737& 344 & 2,\, 4,\, 6\\
\hline
IV&	250,\, 375,\,500&	10& 10	 & 7.5/5.5&	261&	    0.438& 19 &3440&983 & 457 & 2,\, 4,\, 6\\
\hline    
\hline
\end{tabular}
\caption{\label{table} Structual parameters of GMR-GNR detectors ( $\kappa_S = 10$).} 
\end{table*}

Figure~2  shows
the characteristic current  responsivity $R^J$ 
vs  bias voltage $V_G$ calculated  using ~Eq.~(15)
for  the GMR-GNR detectors with different structural parameters listed in Table I. 
As seen from Fig.~2, the characteristic current responsivity exhibits a pronounced maxima at a certain bias voltage.
This voltage,  $V_{Max}$, and the height of the responsivity maximum, max~$R^J$, are determined by the array structural parameters.
 In particular,
for the parameters of  structure I  with $M(2N-1) = 9 - 81$, from Eq.~(16) and Fig.~2 we obtain  the following estimate: max~$R^J \simeq
(0.69 - 6.18)$~ A/W. 
  
\begin{figure}[t]
\centering
\includegraphics[width=8.0cm]{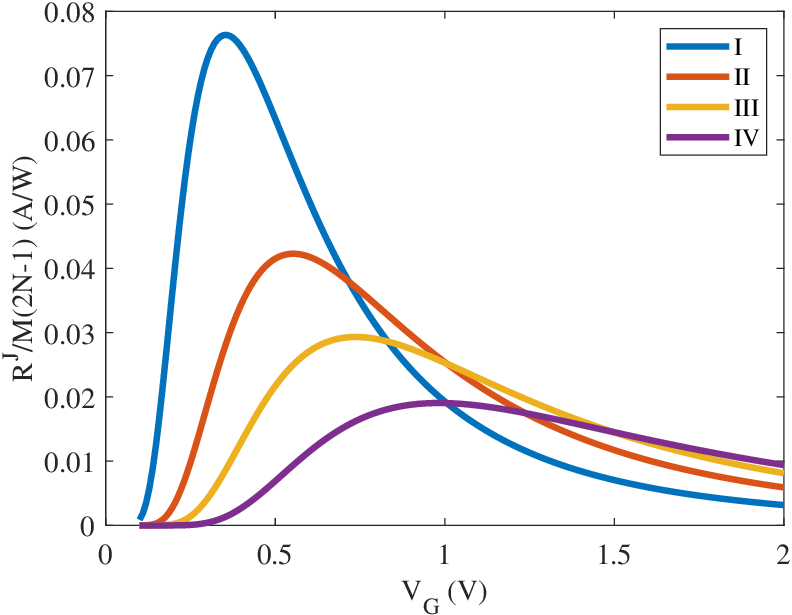}
\caption{Normalized characteristic current responsivity $R^J/[M(2N-1)]$ as a function of  bias voltage $V_G$ calculated  for different parameters corresponding to structures~I - IV.}\label{Fig2}
\end{figure}

 In the GMR-GNR detectors using the rectified voltage as the output,  
 the signal dc voltage $\Delta V_{\omega}$, given by Eq.~(12),
  corresponds to the following formula for the  detector voltage responsivity (in the V/W units)
   $R_{\omega}^V = \Delta V_{\omega}/P_{\omega}$ :

\begin{eqnarray}\label{eq19}
 R_{\omega}^V = 
 R^V\Pi_{\omega}
\end{eqnarray}
with

\begin{eqnarray}\label{eq20}
 R^V = R^J \frac{r_L\rho_{GNR}}{(r_L + \rho_{GNR})}.
\end{eqnarray}
The factor $R^J$ in Eq.~(20) is described by Eq.~(15), but with $V_G$ replaced
by $V_G^V = V_G\,r_{GNR}/(r_L + r_{GNR})$.

If, for example, $r_L = 377$~$\Omega$ (this implies that in reality $r_L \ll r_{GNR}$), 
for the maximum values of the characteristic voltage responsivity  max~$R^V$, corresponding to the above estimate of the current responsivity, we obtain  max~$R^V \simeq (275-3477)$~V/W.

For larger load resistances, in particular,  compared with the GMR-GNR array dc resistance,  $R^V$ can markedly exceed the latter values.
Indeed, setting, for the definiteness, 
 the load resistance $r_L$ to be  equal to the array DC resistance  $r_{GNR}$,
 at the bias voltage $V_{Max}$, corresponding to the maximum of the current responsivity, 
 and using the expressions for $r_{GNR}$ and $\rho_{GNR}$ 
 [see Eqs.~(A11) and (A12)], we obtain

\begin{eqnarray}\label{eq21} 
{\rm max}~ R^V \simeq 
 \frac{7^3\pi^2}{9}\frac{1}{cV_{Tunn}}.
\end{eqnarray}
For the parameters corresponding to structures I - IV, Eq.~(21) yields $R^V \simeq (3.28- 9.11)\times 10^3$~V/W.

\section{Noise equivalent power}

 To evaluate 
 the noise equivalent power, (NEP$_\omega$), of the GMR-GNR detectors receiving the THz radiation with the frequency $\omega$ (in the units W$/\sqrt{{\rm {Hz}}}$),
  we use the following formulas:

  \begin{eqnarray}\label{eq22} 
 {\rm NEP}_{\omega} = \frac{\sqrt{4e{\overline J}}}{ R_{\omega}^J} = \frac{\rm {NEP}}{\Pi_{\omega}},
  \end{eqnarray}
where ${\overline J}$ is the dc current through one GNR in the absence of THz irradiation (dark current), given by Eq.~(A10), and
  
  \begin{eqnarray}\label{eq23} 
{\rm NEP} =\frac{\sqrt{4e{\overline J}}}{ R^J}\nonumber\\
 = 
 \frac{\sqrt{ce}}{4\pi^{3/2}}\sqrt{\frac{c\hbar}{e^2}}
 \biggl(\frac{V_G}{{\overline V}_G}\biggr)^{1/4}\frac{V_G^{7/2}}{V_{Tunn}^2}\exp\biggl(\frac{V_{Tunn}}{2V_G}\biggr)
  \end{eqnarray}
is the characteristic NEP. 

At the bias voltage $V_{Min} = 2V_{Tunn}/15$, corresponding to the NEP minimum , from Eq.~(21) we obtain\\
 $
{\rm min~NEP} \simeq
 \displaystyle\frac{18}{\sqrt{M(2N-1)}}\, {\rm pW}/{\rm Hz}^{1/2}.
$

Figure~3 shows the GMR-GNR detector NEP, calculated using Eq.~(23) for the GMR-GNR detectors with  structures I -IV (i.e., the same structural parameters as in Fig.~2). For the definiteness, we assume that  $M(2N-1)= 9$.
The NEP values of   
 GMR-GNR detectors   exhibit fairly deep   minima as  functions of the bias voltage. The NEP minima are achieved at the bias voltage $V_{Min}$, which is about of  $V_{Max}/2$.

\begin{figure}[t]
\centering
\includegraphics[width=8.0cm]{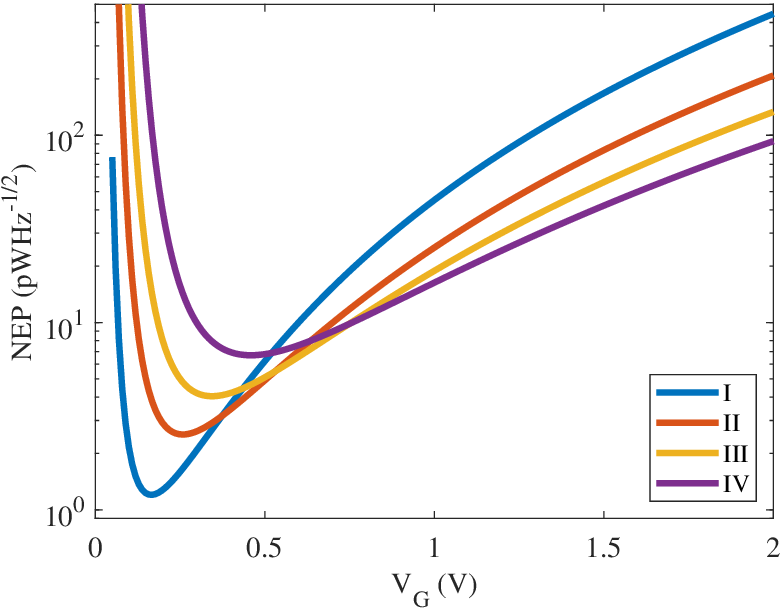}
\caption{Characteristic {\rm NEP}  vs  bias voltage $V_G$ for  structures I - IV with $M(2N-1) = 9$. 
} \label{Fig3}
\end{figure}

\section{Spectral  characteristics of the GMR-GNR detectors} 
 
The excitation of the plasmonic oscillations (standing plasmonic waves with the wave vector directed along the GNRs) by the impinging THz radiation  can result in the resonant response with the responsivities (both the current and voltage responsivities) exhibiting maxima at the plasmonic resonant frequency   $\omega \simeq \Omega/2$  and its harmonics ($\omega \simeq 3\Omega/2, \, 5\Omega/2$,...).
The plasmonic frequency, given by Eq.~(7), is determined by the structural parameters and the bias voltage $V_G$ (via the dependence of $\Omega$ on the carrier Fermi energy $\mu$). 
The main structural parameters, which affect the plasmonic frequency are the 
the width, $2L_G$, of the GMRs and the spacing between them, i.e., the GNR length $2L$). The ratio of this  parameters determines the inter-GMR capacitance $c_G$, and the GMR width, and, therefore  relate
the carrier density in the GNRs (and, hence, the carrier Fermi energy).
Due to the dependence of the plasmonic oscillations on the wave-vector along the GMR,
the length of the latter substantially affects $\Omega$ in line with Eq.~(7).

Figure~4 shows the voltage dependence of the plasmonic frequency calculated for
the structures I - IV with the  different GMR lengths. The plasmonic frequency markedly increases with increasing bias
voltage, although such an increase  is weaker ($\Omega \propto V_G^{1/4}$) than in the standard gated  2D  heterostructure devices (where $\Omega \propto V^{1/2}$. This is common for 2D carrier systems in graphene channels (see, for example,~Refs.~17 and 18). However, a relatively small value of the capacitance 
 in the coplanar conducting arrays
in comparison with the gated structures (like FETs), is a positive factor promoting higher  plasmonic frequencies. As seen from Fig.~4, a decrease in the GMR length $2H$
enables a substantial increase in the plasmonic frequency $\Omega$

\begin{figure}[t]
\centering
\includegraphics[width=8.0cm]{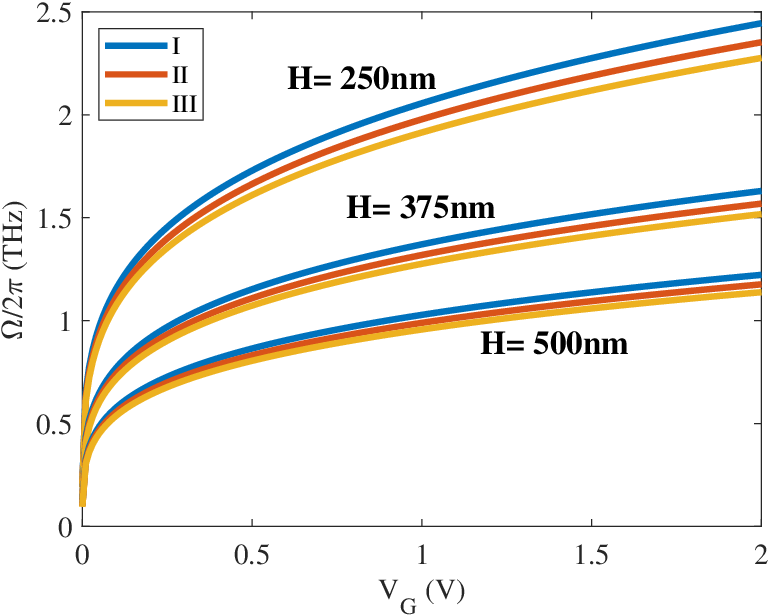}
\caption{Plasmonic frequency $\Omega/2\pi$ vs bias voltage $V_G$
for  structures I - III with different GMR lengths $2H $.
} \label{Fig4}
\end{figure}

\begin{figure*}[t]
\centering
\hspace{+1cm}
\includegraphics[width=16cm]{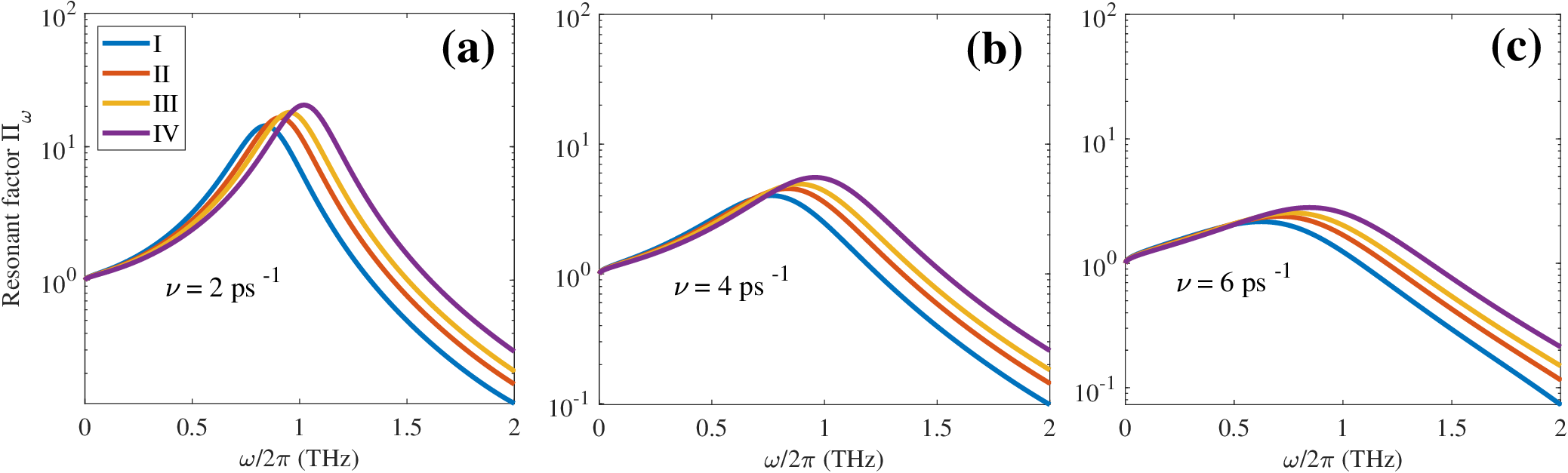}
\caption{
 Plasmonic resonant factor $\Pi_{\omega}$ vs signal frequency $\omega/2\pi$ at  bias voltages $V_G$ corresponding to the maximal current\\ responsivity 
 for  structures I - IV for 
 different carrier collision frequencies $\nu$
 .} 
 \label{Fig5}
\end{figure*}

 Figure~5 shows the spectral dependences of the  plasmonic resonant factor
 $\Pi_{\omega}$
  for the GMR-GNR detector with  structures I -  IV. Calculating   $\Pi_{\omega}$, we accounted for the plasmonic  frequency voltage dependence for the GMR length $2H = 500$~nm..  This resonant factor is sensitive to the plasmonic oscillation quality factor  $Q = 2\Omega/\pi\nu$, which essentially depends on the carrier scattering frequency $\nu$.
 Since    at the plasmonic resonances,  $\Pi_{\omega} \simeq Q^2$, the resonant factor, as seen from Fig.~5,
is large even at the room temperature $\nu$ corresponding to the carriers in graphene on h-BN substrate (see, for example,~Refs.~23 - 31) .  For example,  the carrier mobility in the GMRs, ${\mathcal M} = ev_W^2/\nu$, corresponding to  $\nu = (2 -6)$~ps$^{-1}$ and 
the geometrical parameter of structures I - IV at $V_G = V_{max}$, is in the range $\mathcal M \simeq (1.2 -7.1)\times 10^4$~cm$^2$/V s.
At lower temperatures or in the GMR-GNR arrays with higher plasmonic frequencies (and, hence, operating in the radiation frequencies of several THz), the resonant factor can be particularly large.
 Large values of the resonant factor lead to a substantial increase in the GMR-GNR detector responsivity and
 to the pertinent drop of their NEP.

\section{Comments}

\subsection*{Optimal operation}

The performance of the GMR-GNR detectors under consideration depends on the structural parameters, particularly those determining the barrier height $\Delta$, the bias voltage $V_G$, and the plasmonic resonance quality factor $Q$.  The detector characteristic responsivity $R^J $ large at $V_G \sim V_{Max}$, while the characteristic NEP is small at $V_G \sim V_{Min}$. Hence, the voltage range $V_{\min} \lesssim V_G \lesssim V_{Max}$ can be considered as optimal for the detector operation.  Since $V_{Max}$ and $V_{Min}$ depend on $\Delta$, the optimal operation can be achieved in a certain range of  these quantities' variations.  Figure~6 shows such a range invoking Eq.~(17):
$V_{Max}(\Delta) = (2\pi\,L/7\hbar\,v_We) \Delta^2$ and $V_{Min}(\Delta) = (2\pi\,L/15\hbar\,v_We) \Delta^2$. 
It is assumed, for the definiteness,  that $L= 10$~nm.

\begin{figure}[b]
\centering
\includegraphics[width=8.0cm]{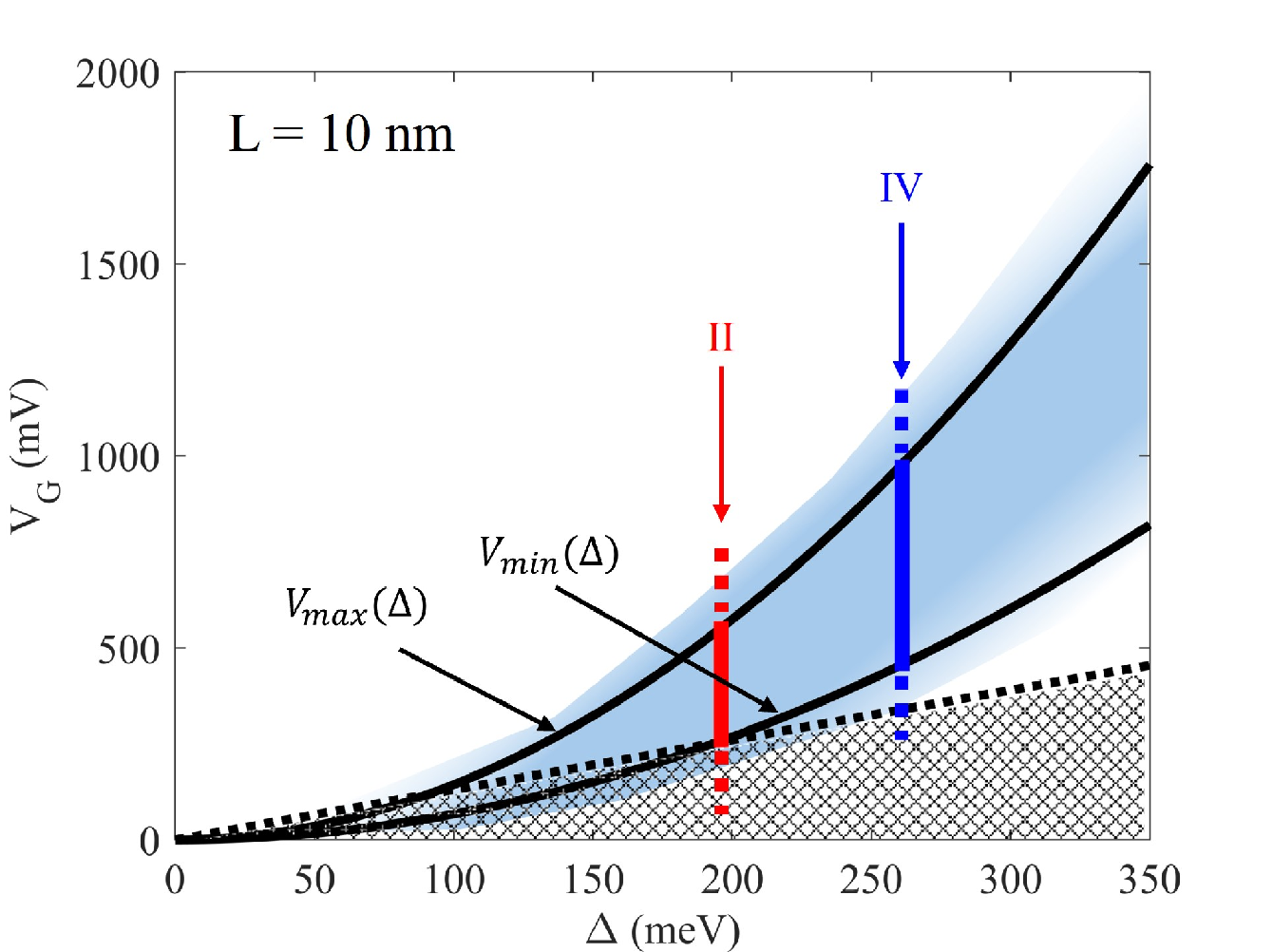}
\caption{Range of the GMR-GNR detector optimal operation (blue-shaded area). The vertical bars correspond to
 structures I and IV. The crosshatched area indicates the range of parameters in which the thermionic dark current can be essential (at $T = 25$~meV).
} \label{FigX}
\end{figure}

\subsection*{ Role of the thermionic current}
The inter-GMR thermionic current  associated with the carriers
overcoming the barrier in  the GNRs was disregarded above. 
Since in the GMR-GNR detectors under consideration, the barrier  height $\Delta$ is virtually insensitive to
the voltage drop across the GNRs (due to the trapezoidal  shape of the barrier), the thermionic processes do not contribute to the ac and rectified current and, hence, do not affect the responsivity, i.e., the plots in Fig.~2.

However,
 the thermionic processes can lead to the following effects:
 First, these processes can contribute to the dark current increasing NEP.
Second, the carrier heating by the impinging THz radiation can  increase the thermionic current(hot-carrier bolometric mechanism of the detector operation).
Both mechanisms  are determined by the ratio $\Delta/T$ and depend on the bias voltage $V_G$.

The tunneling dark current exceeds the  thermionic current (as assumed above)
when $\Delta/T > V_{Tunn}/V_G$,  i.e., according to Eq.(A5), when $V_G > \Delta\,T (\pi\,L/e\hbar\,v_W)$.
As follows from Fig.~6,  the latter conditions are fulfilled for structures I - IV (with the parameters listed in Table I) in wide ranges of $\Delta$ and $V_G$. As for structure I, the thermionic dark current can be comparable
with the tunneling dark current, affecting the value of NEP near its minimum. This implies that the NEP of structure I around its minimum is somewhat underestimated.

 An increase in the carrier effective temperature
stimulated by the THz radiation can result in the bolometric contribution to the output signal.
This is similar to the THz detectors based on the gated graphene structures considered previously.~\cite{7,13,28,32,33}
Our rough  estimate shows that the THz detection associated with the rectification of the inter-GMR current, considered above,
prevails over the hot-carrier bolometric effect if, in particular,   $\Delta/T > V_{Tunn}/V_G +\ln(\tau_{\varepsilon}\nu)T/eV_G)\sim
 [7/2 + \ln(7\tau_{\varepsilon}\nu)T/2eV_{Tunn}]$, where $\tau_{\varepsilon}$ is the energy relaxation time in the GMRs. The latter condition can be stricter than the above one. However, the logarithmic term is about unity even at large values of the product $\tau_{\varepsilon}\nu$. The contribution of the bolometric effect in question can 
be beneficial for the enhancement of the GMR-GNR detectors performance being, however, characterized by a much longer
response time (about of $\tau_{\varepsilon}$).
A more detailed   consideration
of the bolometric effect is beyond the scope of this work and requires a separate study.

\subsection*{ Carrier and plasmon confinement in the GMRs.}

The GMRs in the devices under consideration have a relatively small width $L_G$. This can result in the bandgap opening 
not only in GNRs but in the GMRs as well. However, the former effect is relatively weak because of $L_G \gg w$; it is accounted for by a distinction of the real and effective GMR widths, $w$ and $W$.

The finiteness of the GMR width can lead to the quantization of the plasmon spectrum with the appearance of higher
plasmonic modes. However, due to $H \gg L_G$, such modes are characterized by frequencies much higher than the fundamental plasmonic frequency $\Omega$ and the frequency of the detected THz radiation frequency $\omega$.

\section{Conclusion} 
 
We  proposed the concept of the THz detector based on a coplanar interdigital GMR-GNR array and evaluated its performance.  We showed that such detectors exhibit  high values of room temperature responsivity and  low noise equivalent power, which can provide their strong competitiveness with existing fast  heterostructure THz detectors
(see, for example, Refs.~4, 7, and 34.
The predicted high performance of the GMR-GNR detectors might encourage their fabrication and applications.

\section*{Acknowledgements}
This work  at Research Institute of Electrical Communication (RIEC), Frontier Institute for Interdisciplinary Studies (FRIIS),   and University of Aizu (UoA) was supported by the Japan Society for Promotion of Science (KAKENHI grant  No. 21H04546),
Japan. The work at Rensselaer Polytechnic Institute (RPI) was supported   by the AFOSR (contract No. FA9550-19-1-0355).

\section*{Author declaration}
\subsection*{Conflict of interest}
The authors have no conflict to disclose.

\section*{Data availability}

The data that support the findings of this study are available within the article.

\section*{Appendix A. Electron-hole tunneling currents through GNRs }
 \setcounter{equation}{0}
\renewcommand{\theequation} {A\arabic{equation}}

We consider the GMR-GNR arrays with
 the trapezoidal energy barriers for the electrons and holes in the  GMRs (in the absence of the bias voltage), i.e., with a rather short transition area length.
 In such a case, the top of the barrier lowering is a weak function of the inter-GMR voltage. Hence, the variation of the inter-GMR current in the GNRs with varying voltage is of tunneling origin.
Following the Landauer-Buttiker formula~\cite{35,36} applied to the 1D electron and hole transport through the GNRs and accounting for the expression for the barrier transparency,  the tunneling current can be presented as

\begin{eqnarray}\label{eqA1}
J = \frac{4e}{\pi\hbar}\int_0^{\infty}
d\varepsilon \frac{T(\varepsilon)}{\displaystyle\biggl(\exp\frac{\varepsilon -\mu}{T}\biggr) +1},
\end{eqnarray}
where at $eV_G \geq \Delta$ when the trapezoidal barrier transforms into  the triangular one, the tunneling transparency $T(\varepsilon)$ reads (disregarding possible short transition regions near  the GMR-GNR contacts, in which the GNR width varies):

\begin{eqnarray}\label{eqA2}
T(\varepsilon) \simeq \exp\biggl[ - \frac{2\Delta}{\hbar\,v_W} \int_0^{(\Delta-\varepsilon)/eE}dx \sqrt{1 -\biggl(\frac{\varepsilon+ eEx}{\Delta}\biggr)^2}\biggr]\nonumber\\
\simeq \exp\biggl[ - \frac{\pi\Delta^2}{2\hbar\,v_WeE}\cdot K\biggl(\frac{\varepsilon}{\Delta}\biggr)\biggr]
\end{eqnarray}
is the GNR barrier tunneling transparency with

$$
K(\delta) =\frac{
4}{\pi}\int_0^{1-\delta}d\zeta\sqrt{1 - \zeta^2}
$$
and $E=V/2L$ being the electric field along the GNRs and $\Delta$ is the barrier height for the 
carriers in the GNRs.
The carrier energy spectra in the GNRs and the GMRs  are  
$\varepsilon^{\pm} = \pm\sqrt{\Delta_{GNR}^2 + (pv_W)^2}$ and 
$\varepsilon^{\pm} = \pm\sqrt{\Delta_{GMR}^2 + (pv_W)^2}$ , 
where the pertinent bandgap openings are estimated as  $2\Delta_{GNR} \simeq \pi\,\hbar\,v_W/w$ and  $2\Delta_{GMR} \simeq \pi\,\hbar\,v_W/L_G$, 
and $p$ is the momentum along  GNRs and GMRs. Hence,   $\Delta = (\pi\hbar\,v_w/W)$ with $W = w/(1 - w/2L_G)$.

If $\varepsilon \sim \mu \ll \Delta$,
$K(\delta) \simeq 1 -(4/\sqrt{2}\pi)\delta^{3/2} \lesssim  1$, at $T \ll \mu$.
In this case,
\begin{eqnarray}\label{eqA3}
T(\varepsilon) \simeq\exp\biggl( - \frac{V_{Tunn}}{V_G}\biggr),
\end{eqnarray}
so that 
Eqs.~(A1) - (A3) yield

\begin{eqnarray}\label{eqA4}
J = \frac{4e}{\pi\hbar}\int_0^{\infty}
d\varepsilon \frac{\displaystyle\exp\biggl[ - \frac{\pi\Delta^2}{2\hbar\,v_WeE} \cdot
K\biggl(\frac{\varepsilon}{\Delta}\biggr)\biggr]}{\displaystyle\exp\biggl(\frac{\varepsilon -\mu}{T}\biggr) +1}\nonumber\\
\simeq\frac{4e^2\sqrt{{\overline V}_GV}}{\pi\hbar}\exp\biggl( - \frac{V_{Tunn}}{V}\biggr). 
\end{eqnarray}
Here

\begin{eqnarray}\label{eqA5}
{\overline V}_G =\frac{\pi\,c_G\hbar^2v_W^2}{2e^3L_G}
\end{eqnarray}
and

\begin{eqnarray}\label{eqA6}
V_{Tunn} = \frac{\pi\Delta^2L}{\hbar\,v_We}= \frac{\pi^3\hbar\,v_WL}{eW^2}.
\end{eqnarray}
is
the characteristic tunneling voltage.

 Equation~(A4) leads to

\begin{eqnarray}\label{eqA7}
\sigma_{GNR} =  \frac{d J}{dV}\biggr|_{V_G}
\simeq  \overline{j}\frac{V_{Tunn}}{V_G}\biggl(\frac{V_{Tunn}}{V_G}+ \frac{1}{2}\biggr)\nonumber\\
=\frac{4e^2}{\pi\hbar}\sqrt{\frac{{\overline V}_G}{V_G}}\biggl(\frac{V_{Tunn}}{V_G}+ \frac{1}{2}\biggr)\exp\biggl( - \frac{V_{Tunn}}{V_G}\biggr)\nonumber\\
\simeq \frac{4e^2}{\pi\hbar}\sqrt{\frac{{\overline V}_G}{V_G}}\frac{V_{Tunn}}{V_G}\exp\biggl( - \frac{V_{Tunn}}{V_G}\biggr),
\end{eqnarray}  

\begin{eqnarray}\label{eqA8}
\eta_{GNR} =  \frac{1}{2}\frac{d^2 J}{dV^2}\biggr|_{V_G}
\simeq  \frac{\overline{j}}{2V_G^2}
\biggl(\frac{V_{Tunn}^2}{V_G^2}-\frac{1}{2}\frac{V_{Tunn}}{V_G} -\frac{1}{4}\biggr)\nonumber\\
= \frac{2e^2}{\pi\hbar}
\frac{\sqrt{{\overline V}_GV_G}}{V_G^2}
\biggl(\frac{V_{Tunn}^2}{V_G^2}-\frac{1}{2}\frac{V_{Tunn}}{V_G} -\frac{1}{4}\biggr)
\exp\biggl( - \frac{V_{Tunn}}{V_G}\biggr)\nonumber\\
\simeq\frac{2e^2}{\pi\hbar}
\sqrt{\frac{{\overline V}_G}{V_G}}
\frac{V_{Tunn}^2}{V_G^{3}}
\exp\biggl( - \frac{V_{Tunn}}{V_G}\biggr),\,
\end{eqnarray}  
 so that 
   
\begin{eqnarray}\label{eqA9}
\frac{\eta_{GNR}}{\sigma_{GNR}} \simeq  \frac{1}{2V_G}\frac{V_{tunn}}{V_G}.
\end{eqnarray}  
Here
 
\begin{eqnarray}\label{eqA10}
{\overline J}\simeq\frac{4e^2\sqrt{{\overline V}_GV_G}}{\pi\hbar}\exp\biggl( - \frac{V_{Tunn}}{V_G}\biggr) 
\end{eqnarray}
is  the DC current  in the absence of THz irradiation (dark current per one GNR). 
A simplification of Eqs.~(A7) - (A9) is justified when $V_G$ is markedly smaller than $V_{Tunn}$. The latter corresponds to reality.
The above equations are valid if $eV_G \geq \Delta$, i.e., when the barrier becomes triangular under the bias voltage.

In particular, Eq.~(A10) corresponds to the GMR-GNR array differential resistance and  DC resistance (which are different) equal to

\begin{eqnarray}\label{eq11} 
\rho_{GNR} = \frac{\pi\hbar}{4e^2M(2N-1)}  \sqrt{\frac{ V_G}{{\overline V}_G}}
\frac{V_G}{V_{Tunn}}\exp\biggl(\frac{V_{tunn}}{ V_G}\biggr)
\end{eqnarray}
and

\begin{eqnarray}\label{eqA12}
r_{GNR} = \frac{V_G}{M(2N-1)\,{\overline J}}\nonumber\\
 \simeq\frac{\pi \hbar}{4M(2N-1)\,e^2}\sqrt{\frac{V_G}{\overline V}_G}\exp\biggl( \frac{V_{Tunn}}{V_G}\biggr), 
\end{eqnarray}
respectively, therefore
 
\begin{eqnarray}\label{eqA13}
\frac{\rho_{GNR}}{r_{GNR}}\simeq\frac{V_G}{V_{tunn}}. 
\end{eqnarray}
In particular, for structure II at $V_G = V_{Max} $ we find $r_{GNR} \simeq [340/M{2N-1}]$~k$\Omega$.
 %

\end{document}